\documentclass[12pt]{article}
\usepackage{enumerate,eepic,epic,amssymb,amsfonts,times,color,epsfig,
fancybox,graphicx,amsmath,pifont,pxfonts,txfonts,bbm}
\makeatletter
\ifcase\@ptsize
 \font\teneufm=eufm10
 \font\seveneufm=eufm7
 \font\fiveeufm=eufm5
 \font\teneusm=eusm10
 \font\seveneusm=eusm7
 \font\fiveeusm=eusm5
\or
 \font\teneufm=eufm10 scaled \magstephalf
 \font\seveneufm=eufm7
 \font\fiveeufm=eufm5
 \font\teneusm=eusm10 scaled \magstephalf
 \font\seveneusm=eusm7
 \font\fiveeusm=eusm5
\or
 \font\teneufm=eufm10 scaled \magstep1
 \font\seveneufm=eufm7
 \font\fiveeufm=eufm5
 \font\teneusm=eusm10 scaled \magstep1
 \font\seveneusm=eusm7
 \font\fiveeusm=eusm5
\fi

\newfam\eufmfam
\newfam\eusmfam
\textfont\eufmfam=\teneufm  \scriptfont\eufmfam=\seveneufm
  \scriptscriptfont\eufmfam=\fiveeufm
\textfont\eusmfam=\teneusm  \scriptfont\eusmfam=\seveneusm
  \scriptscriptfont\eusmfam=\fiveeusm

\def\frak{\ifmmode\let\next\frak@\else
 \def\next{\errmessage{Use \string\frak\space only in math mode}}\fi\next}
\def\frak@#1{{\frak@@{#1}}}
\def\frak@@#1{\fam\eufmfam#1}

\def\sh{\ifmmode\let\next\sh@\else
 \def\next{\errmessage{Use \string\sh\space only in math mode}}\fi\next}
\def\sh@#1{{\sh@@{#1}}}
\def\sh@@#1{\fam\eusmfam#1}

\ifcase\@ptsize
 \font\tenmsa=msam10
 \font\sevenmsa=msam7
 \font\fivemsa=msam5
 \font\tenmsb=msbm10
 \font\sevenmsb=msbm7
 \font\fivemsb=msbm5
\or
 \font\tenmsa=msam10 scaled \magstephalf
 \font\sevenmsa=msam7
 \font\fivemsa=msam5
 \font\tenmsb=msbm10 scaled \magstephalf
 \font\sevenmsb=msbm7
 \font\fivemsb=msbm5
\or
 \font\tenmsa=msam10 scaled \magstep1
 \font\sevenmsa=msam7
 \font\fivemsa=msam5
 \font\tenmsb=msbm10 scaled \magstep1
 \font\sevenmsb=msbm7
 \font\fivemsb=msbm5
\fi

\newfam\msafam
\newfam\msbfam
\textfont\msafam=\tenmsa  \scriptfont\msafam=\sevenmsa
  \scriptscriptfont\msafam=\fivemsa
\textfont\msbfam=\tenmsb  \scriptfont\msbfam=\sevenmsb
  \scriptscriptfont\msbfam=\fivemsb

\def\Bbb{\ifmmode\let\next\Bbb@\else
 \def\next{\errmessage{Use \string\Bbb\space only in math mode}}\fi\next}
\def\Bbb@#1{{\Bbb@@{#1}}}
\def\Bbb@@#1{\fam\msbfam#1}
\def\hexnumber@#1{\ifnum#1<10 \number#1\else
 \ifnum#1=10 A\else\ifnum#1=11 B\else\ifnum#1=12 C\else
 \ifnum#1=13 D\else\ifnum#1=14 E\else\ifnum#1=15 F\fi\fi\fi\fi\fi\fi\fi}
\def\msa@{\hexnumber@\msafam}
\def\msb@{\hexnumber@\msbfam}
\mathchardef\square="0\msa@03

\makeatother
\newcommand{\beq}{\begin{equation}}
\newcommand{\eeq}{\end{equation}}
\newcommand{\ba}{\begin{array}}
\newcommand{\ea}{\end{array}}
\newcommand{\bea}{\begin{eqnarray}}
\newcommand{\eea}{\end{eqnarray}}
\newcommand{\bean}{\begin{eqnarray*}}
\newcommand{\eean}{\end{eqnarray*}}

\newtheorem{theorem}{Theorem}[section]
\newtheorem{prop}[theorem]{Proposition}
\newtheorem{lem}[theorem]{Lemma}
\newtheorem{defi}[theorem]{Definition}

\newtheorem{remark}[theorem]{Remark}

\newtheorem{proof}{Proof.}
\newenvironment{pf}{\begin{proof} \rm}{\hfill$\square$ \end{proof}}

\makeatletter
\@addtoreset{equation}{section}

\makeatother

\newcommand{\rref}[1]{(\ref{#1})} 

\def\endpf{\begin{flushright}$\square$\end{flushright}}

\begin{document}
\begin{titlepage}
\begin{center}
{\huge 
A new construction of}\vskip 0.5truecm \noindent  {\huge the Drinfeld--Sokolov Hierarchies}
\end{center}
\vspace{0.8truecm}
\begin{center}
{\large
Paolo Casati, 
\vskip  0.8truecm Dipartimento di Matematica e applicazioni\\
 Universit\`a di Milano-Bicocca\\
Via Cozzi  55, I-20125 Milano, Italy}
 
\end{center}
E--mail:  paolo.casati@unimib.it\\
\vspace{0.2truecm}

\abstract{\noindent
The Drinfeld--Sokolov hierarchies  are integrable hierarchies associated  with every affine Lie algebra. We present a new construction of such hierarchies, which only requires the computations of a formal Laurent series.}\vskip 1truecm\noindent{{\bf Key Words}:Integrable Hierarchies of PDE, Lie algebras,
Irreducible Representations,  Drinfeld--Sokolov construction.} 
\vskip 1truecm\noindent{\bf Mathematics Subjects Classification:} Primary 37K10 Secondary 53D5\vskip 2truecm\noindent

\end{titlepage}
\section{Introduction}
The discovery made by Drinfeld and Sokolov in their celebrated  paper \cite{DS}, that to any affine Kac--Moody Lie algebra and to any  vertex of its (extended)  Dynkin diagram corresponds a hierarchy of completely integrable nonlinear partial differential equations,   is a hallmark in  the theory of integrable systems.
These hierarchies (usually called Drinfeld--Sokolov hierarchies)  have  been further and intensively studied in the last twenty years by many mathematicians   with the help of various techniques;  \cite{CP} \cite{BdGHM}  \cite{BdGHM1} \cite{CFMP2} \cite{CFMP3}  \cite{CFMPtr} \cite{CFMPfract} \cite{FHM} \cite{DF} \cite{FM} \cite{F} \cite{DFG} 
 \cite{KM} \cite{CDO} \cite{W} \cite{DLZ} \cite{LWZ} (in particular see \cite{DLZ} for a very   beautiful and recent paper on this subject). \par
The aim of this paper is to provide a new simple and algorithmic way to construct the equations and  the conserved quantities of the Drinfeld and Sokolov hierarchies corresponding to the untwisted affine Kac--Moody Lie algebras $A^{(1)}_n$,   $B^{(1)}_n$ $C^{(1)}_n$ $D^{(1)}_n$  
and the vertex $c_0$ of the Dynkin diagram  (i.e., the one added to the Dynkin
diagram of the associated simple Lie algebra). \par 
Such construction may be briefly described as follows. Let  $h(z)=z+\sum_{i>0}h_iz_i$ be a formal Laurent series in $z$ whose   coefficients $h_i$ are smooth functions from the unit circle ${S}^1$ 
in $\mathbbm{C}$: $h_i\in C^\infty(S^1,\mathbbm{C})$ and  define its Fa\`a di Bruno iterates as $h^{(k)}=(\partial_x  +h(x))^k(1)$ where $\partial_x$ is the derivative along the coordinate $x$  on ${S}^1$ (thus  $h^{(0)}=1$, $h^{(1)}=h(x)$, $h^{(2)}=h_x+h^2$,...). Then  consider the following constrains  in terms of ODE's of Riccati type on the Laurent series $h(x)$:\par\noindent 
\begin{tabular}{lll}
&$
z^{n+1}=h^{(n+1)}-\sum_{i=0}^{n-1}u_kh^{(i)}
$ &
for the Lie algebra $A^{(1)}_n$.\\
&$ z^{2n}h=h^{(2n+1)}-\sum_{k=0}^{n-1}u_kh^{(2k+1)}-\sum_{k=0}^{n-1}(\partial_x+h)^{(2k+1)}(u_k)$
& for the Lie algebra $B^{(1)}_n$.
\\
&$z^{2n}h=h^{(2n)}-\sum_{k=0}^{n-1}u_kh^{(2k)}-\sum_{k=0}^{n-1}(\partial_x+h)^{(2k)}(u_k)$  &
for the Lie algebra $C^{(1)}_n$.
\\
&$z^{2n-2}h=h^{(2n-1)}-\sum_{k=1}^{n-1}u_kh^{(2k-1)}-\sum_{k=0}^{n-1}(\partial_x+h)^{(2k-1)}(u_k) $\\&$\phantom{z^{2n-2}h=}- u_0h^{(-1)}_{u_0}$ 
&   for the Lie algebra $D^{(1)}_n$
\end{tabular}\vskip 0.5truecm \noindent
where the Laurent series $h^{(-1)}_{u_0}$ is uniquely defined by the relation $(\partial_x+h)( h^{(-1)}_{u_0})=u_0$.\par 
The coefficients of the formal Laurent series, which solve such equations, 
can be computed recursively in a pure algebraic  way (i.e., without performing any integrations) as   differential polynomials in the  free fields $u_i$.\par\noindent 
The above  constructed Fa\`a di Bruno polynomials define on
the space $\mathcal{L}$   of all Laurent series which are truncated from above i. e.,  the space
$$
{\mathcal{L}}:=\{\sum_{-\infty}^N l_i z^{-i}\vert\ l_i\in \mathbbm{C},\  N 
\in {\Bbb Z}\} 
$$
the splitting 
$$
\mathcal{L}=\mathcal{L}_+\oplus \mathcal{L}_-=\mbox{span}\{h^{(k)}\vert\ k\geq  0\}\oplus {\mbox{span}\{z^k\vert\ k< 0\}}.
$$
Let  $\pi_+$ be  the corresponding  projection of $\mathcal{L}$ onto $\mathcal{L}_+$ and 
let $\mathcal{N}_{\widehat{\mathfrak{g}}}$ be the subset of $\mathbbm{N}$ equal to $\{i\in \mathbbm{N}\vert \ i\neq 0\mbox{mod}(n+1)\}$ if the considered Kac--Moody algebra ${\widehat{\mathfrak{g}}}$ is of type $A_1^{(n)}$ and $\{2i+1\vert i\in  \mathbbm{N}\}$ if is ot type $B$,$C$, or $D$.
Define  $H^{(i)}\in \mathcal{L}$ as 
$$
  H^{(k)}=\pi^+(z^k)\qquad k\in \mathcal{N}_{\widehat{\mathfrak{g}}}.
$$
Then the  equations of the corresponding hierarchies of integrable PDE will have the form
$$
\frac{\partial h}{\partial t_k}=\partial_x H^{(k)}\qquad k\in \mathcal{N}_{\widehat{\mathfrak{g}}}.
$$
To achieve such results we shall mainly use an approach to the Drinfeld--Sokolov reduction based on a  bihamiltonian reduction theorem \cite{CMP1}
\cite{CP}, \cite{CDO}  and the theory of simple and affine  Lie algebras \cite{K}.\par
The paper is organized as follows. In the second section, divided in two subsections, we briefly review the bihamiltonian theory of the integrable system, recall the  bihamiltonian reduction theorem, and  define a bihamiltonian 
structure on the affine Kac--Moody Lie algebras  describing  its property. In the third  and final section, divided in three subsections, we  
finally described into details the construction of the Drinfeld--Sokolov hierarchies outlined above.\par
The author wishes to thank Andriy Panasyuk for many usefull discussions about the
geoemtry of the  bihamiltonian manifolds, and professor Laszlo Feher, for appreciating this work and for drawing my attention to the papers
\cite{FHM} \cite{DF} \cite{FM} \cite{F} \cite{DFG}.
\section{Preliminary results}
In this section we   recall  some background material mainly concerning   the bihamiltonian theory of  the integrable systems and the affine Kac--Moody Lie algebras and their bihamiltonian structures. Detailed descriptions of these topics  may be found in the papers \cite{CMP1} \cite{CP} \cite{CFMPfract} \cite{CFMP2} \cite{DLZ}  \cite{P} or in the books \cite{D} \cite{B}.\par

\subsection{The bihamiltonian theory of the integrable systems} 
A Poisson manifold is a smooth manifold $\mathcal  M$ endowed with a Poisson bracket $\{\cdot,\cdot\}$ i.e, a bilinear skew symmetric composition law of 
$C^\infty(\mathcal{M},\mathbbm{C})$ fulfilling the Leibniz rule and the Jacobi identity. The corresponding Poisson tensor
$P$ is the bivector tensor field on $\mathcal M$, viewed as a linear skew symmetric map between the cotangent and the tangent space bundle : $P:T^*\mathcal{M}\to 
T\mathcal{M}$, defined at any point $m\in \mathcal{M}$ by  
$$
\{F,G\}(m)=\langle dF,P_mdG\rangle_m\qquad \forall F,G \in C^\infty(\mathcal{M},\mathbbm{C})
$$
where $\langle\cdot ,\cdot \rangle_m$ is the pairing between $T^*_m\mathcal{M}$ and $T_m\mathcal{M}$.\par
A bihamiltonian manifold $\mathcal  M$ is a smooth manifold equipped with two compatible Poisson brackets, i.e., two Poisson 
brackets $\{\cdot,\cdot\}_0$  and $\{\cdot,\cdot\}_1$ such that the pencil $$
\{\cdot,\cdot\}_\lambda=\{\cdot,\cdot\}_1-\lambda\{\cdot,\cdot\}_0
$$ 
is a Poisson bracket  for any  $\lambda \in\mathbbm{C}$. We denote by $P_\lambda$ the corresponding pencil of Poisson tensors.
A bihamiltonian vector field $X$ on $\mathcal M$ is a vector field which is Hamiltonian with respect to both Poisson brackets (and therefore with respect to any Poisson bracket $\{\cdot,\cdot\}_\lambda$).\par 
The central idea of the bihamiltonian theory of the hierarchies of PDE's is to view them as  collections of bihamiltonian vector fields on a (usually infinite dimensional) bihamiltonian manifold $\mathcal M$.\par
There are two classical strategy  to  construct such integrable systems. The first one assumes the  
nondegeneracy of one of
the Poisson tensor associated to the two compatible Poisson brackets  
and uses the so--called recursion operator. The second one, which will be used in this paper, uses the Casimir functions   of the (Poisson)  pencil.
More precisely we shall look  for formal Laurent series $H(\lambda)=\sum_{k\geq -N} H_k\lambda^{-k}$, $H_k\in C^\infty(\mathcal{M},\mathbbm{C})$  which are Casimirs of the Poisson pencil:
$$
\{F,H(\lambda)\}_\lambda=0\qquad \forall F\in  C^\infty(\mathcal{M},\mathbbm{C}).
$$
In fact developing this equation in powers of $\lambda$  we obtain a hierarchy $\{ X_k \}_{k\geq N }$ of bihamiltonian vector fields:
$$
\begin{array}{lll}
&\{F,H_{N}\}_0=0&\\
&X_k(F)=\{F,H_{k-1}\}_0=\{F,H_{k}\}_1&\qquad \forall F\in  C^\infty(\mathcal{M},\mathbbm{C})\quad k\leq N.
\end{array}
$$
In this paper we  shall  tackle this very hard problem in the contest of bihamiltonian brackets defined on affine Kac--Moody Lie algebras, where it can be solved by using the generalization of the dressing method of Zakharov and Shabat proposed by Drinfeld and Sokolov \cite{DS}. \par
Actually the bihamiltonian hierarchies, which are interesting in,  are not directly defined on such algebras but rather on reduced bihamiltonian manifolds, constructing by means of  the 
\begin{theorem}\label{BHR} (\cite{CMP1} Prop 1.1). Let $(\mathcal{M},\{\cdot,\cdot\}_0, \{\cdot,\cdot\}_1)$ be a bihamiltonian manifold,  $\cal S$  a symplectic submanifold  of $P_0$, $D$ and $E$  the distributions
$D=P_1 \mbox{Ker}(P_0)$,  $E=T{\cal S}\cap D$.   Then  the distribution  $E$  is integrable and, if the   quotient
space $\mathcal N={\mathcal S}/E$ is a manifold, it is a bihamiltonian manifold 
endowed with the reduced  Poisson pencil $\{\cdot,\cdot\}^{\mathcal{N}}_\lambda$ defined uniquely by the relation 
$$\{f,g\}^{\mathcal{N}}_\lambda\circ\pi=\{F,G\}^{\mathcal{M}}_\lambda\circ i\qquad \forall f,g\in C^\infty(N,\mathbbm{C})
$$
where  $i$ and $\pi$ are the canonical injection of $\mathcal{S}$ in  $\mathcal{M}$ and the canonical projection of 
$\mathcal{M}$ onto $\mathcal{N}$ respectively, and  $F$ and $G$ are any pair of smooth functions, which extend  the functions  $f$
and $g$ of $\mathcal{N}$ into $\mathcal{M}$, and are constant on $D$ (i.e., $F\circ i=f\circ \pi$ and $\{F,K\}_1=0$ for any Casimir $K$ of $P_0$).  \end{theorem} 
\subsection{Affine  Lie algebras as Bihamiltonian Manifolds}
Let $\mathfrak{g}$ be a simple Lie algebra over $\mathbbm{C}$, $G$ the corresponding connected and simply connected Lie
group. Fix a nondegenerate symmetric invariant bilinear form  $\langle\cdot,\cdot\rangle_\mathfrak{g}$  on $\mathfrak{g}$. The associated (non twisted) affine Lie algebra $\widehat{\mathfrak{g}}$ is a semidirect product of the central extensions of the  loop algebra $L({\mathfrak{g}})=C^\infty(S^1,\mathfrak g)$  and a derivation $d$, more precisely 
\begin{defi}\label{afkcl} The affine (non twisted) Lie algebra 
$\widehat{\mathfrak{g}}$ associated with the finite dimensional simple Lie 
algebra $\mathfrak{g}$ is the complex vector space
$$
\widehat{\mathfrak{g}}=C^\infty(S^1,\mathfrak g)
\oplus{\Bbb C}K\oplus{\Bbb C}d
$$
endowed with the following Lie bracket:
\beq
\label{afkmlb}
\begin{array}{ll}
&\left[P(x)+\mu_1K +\nu_1d,Q(x)+\mu_2K +\nu_2d
\right]_{\widehat{\mathfrak{g}}}\\
&=\left[P(x),Q(x)\right]+\nu_1 x\frac{dP(x)}{dx}-\nu_2x
\frac{dQ(x)}{dx}+ \left(\int_{S^1}\langle\frac{P(x)}{dx},Q(x)\rangle_\mathfrak{g}\right) K
\end{array}
\eeq
where $[\cdot,\cdot]$ is the bracket 
in the loop Lie algebra $L({\mathfrak{g}})=C^\infty(S^1,\mathfrak g)$.\end{defi} 
In what follows the derivation $d$ will not play any role.\par\noindent
Let  ${\widehat{\mathfrak{g}}}^{\ *}$ be the space of linear functionals on $\widehat{\mathfrak{g}}$ of the following form
\beq
\label{notdeg}
\mathcal{L}_{(P(x)+aK)}\left(Q(x)+bK\right)=\langle (P(x)+aK),Q(x)+bK\rangle=\int_{S^1} \langle P(x),Q(x)\rangle_\mathfrak{g} dx+ab.
\eeq
We identify ${\widehat{\mathfrak{g}}}^{\ *}$ with $\widehat{\mathfrak{g}}$. Being  $\mathcal{M}=\widehat{\mathfrak{g}}$ a flat manifold we may further identify the tangent space at any point with $\mathcal{M}$ itself. Using these identifications it is easy to compute the  canonical 
Lie Poisson tensor of $\widehat{\mathfrak{g}}$,     as
\beq
\label{canPt}
P_{(S,K)}(V)=K\partial_xV+\left[S,V\right].
\eeq
It can be easily shown \cite{Z}  that this Poisson tensor is compatible with the constant Poisson tensors obtained by freezing it  in any point of $\widehat{\mathfrak{g}}$.\par
In order to specify the choice  which will lead to  the Drinfeld Sokolov hierarchies we must briefly describe 
the structure of the simple (finite dimensional) Lie algebra $\mathfrak{g}$.\par 
Let $\mathfrak{h}$ be  a Cartan subalgebra of $\mathfrak{g}$, and let  $\Delta$ be   the corresponding root system.
Let $\Pi=\{\alpha_1, \dots,\alpha_n\}$  be  a set of simple roots in  $\Delta$ 
(where $n$ is the rank of $\mathfrak{g}$), and $\Delta=\Delta^+\cup\Delta^-$
the associated  decomposition of $\Delta$ in positive (resp. negative) roots, then we have the Cartan decomposition
$$
\label{Cardec}
\mathfrak{g}=\mathfrak{n^+}\oplus\mathfrak{h}\oplus\mathfrak{n}^-=
\left(\bigoplus_{\alpha\in \Delta^+}\mathfrak{g}_\alpha \right)\oplus\mathfrak{h}\oplus\left(\bigoplus_{\alpha\in \Delta^+}\mathfrak{g}_{-\alpha}\right)
$$
where $\mathfrak{g}_\alpha = \{X\in \mathfrak{g}\vert\ \left[H,X\right]=\alpha(H)X\ \forall H\in \mathfrak{h}\}$ 
is the root  space of $\alpha$.  Let, for any root $\alpha\in \Delta$, $H_\alpha$ be the 
corresponding coroot with respect to the bilinear form $\langle\cdot,\cdot\rangle_\mathfrak{g}$, and for any $\alpha\in \Delta^+$ let $X_\alpha$  be a basis of $\mathfrak{g}_\alpha$  and let $Y_{\alpha}$  in  $\mathfrak{g}_{-\alpha}$  be defined by the requirement $\left[X_\alpha,Y_{\alpha}\right]= H_\alpha$.
Let $\theta$ be  the maximal root of $\Delta$ with respect the 
 above defined decomposition (i.e, $\theta+\alpha_i\notin \Delta$ for all $\alpha_i\in \Pi$); and denote by $A$ 
the constant function of $C^\infty(S^1,\mathfrak{g})$ whose value is $Y_\theta$.\par
The hierarchies of Drinfeld and Sokolov are  
 bihamiltonian with respect to the reduction  of the bihamiltonian pair $P_1$, $P_0$ where $P_1$ is the canonical Poisson tensor
\rref{canPt} and $P_0$ is the constant Poisson tensor  
\beq
\label{costP}
(P_0)_{(S,K)}(V)=\left[A,V\right].
\eeq
To perform the bihamiltonian reduction we have only to choose the appropriate symplectic leaf of $P_0$. Following Drinfeld and Sokolov let us choose that passing through the point 
$$
B = \sum_{\beta\in \Pi}X_{\beta}+K.
$$
(In what follows we normalize the value of the central extension to $-1$). In this setting the integrable distribution $E$ can be characterized by the 
\begin{theorem}\label{prop1F} \cite{CP} The subspace ${\mathfrak{g}}_{AB}:=\{V\in {\mathfrak{g}}_{A}\vert V_x+\left[V,B\right]
\in{\mathfrak{g}}_{A}^\perp\}$
is a subalgebra of $\mathfrak{g}$ contained in the nilpotent  subalgebra  of loops with values in the maximal 
nilpotent subalgebra ${\mathfrak {n}}^-=\oplus_{\alpha\in \Delta^-}\mathfrak{g}_\alpha$. Therefore the corresponding group
$G_{AB}=\exp({\mathfrak{g}}_{AB})$ is well defined. The distribution $E$ is spanned by the vector fields 
$(P_1)_B(V)$ with $V$ belonging to ${\mathfrak{g}}_{AB}$, and its integral leaves are the orbits of the 
gauge action of $G_{AB}$ on $\mathcal{S}$ defined by:
\beq
\label{gauact}
S'=JSJ^{-1}+J_xJ^{-1}.
\eeq  
\end{theorem}
The characterization of the chosen symplectic leaf $\mathcal{S}$ given in the previous Theorem plays a pivotal role in the present work. In particular  it allows us to compute a submanifold of $\mathcal S$ transversal to the distribution $E$.
\begin{defi}\label{smtra} A transversal submanifold to the distribution $E$ is a submanifold $\mathcal{Q}$ of  $\mathcal{S}$, which intersects
every integral leaves of the distribution $E$ and therefore the orbit of the group action in one and only one point. This condition implies the following relations on the tangent space:
\beq
\label{trans}
T_{q}\mathcal{S}=T_{q}\mathcal{Q}\oplus E_q\qquad \forall q\in \mathcal{Q}
\eeq
\end{defi}
\section{The Drinfeld Sokolov  Hierarchies}
This section is devoted to  present the main result of this paper: the new algorithmic construction of the Drinfeld--Sokolov hierarchies outlined in the introduction.
\subsection{ The Bihamiltonian setting of the Drinfeld Sokolov hierarchies }
As we have already seen,  we need  to  find    Casimirs of the Poisson pencil $P_\lambda=P_1-\lambda P_0$. More precisely we shall look for solutions of  the  Casimir's equation
\beq
\label{Caseq}
P_\lambda(V(\lambda))=V_x+\left[V,S+\lambda A\right]=0\qquad S\in \mathcal{S}
\eeq 
which are formal Laurent series $V(\lambda)=\sum_{k=-m}^\infty V_k\lambda^{-k}$,  
$m\in \mathbbm{Z}$ 
whose coefficients are  one forms defined at least on the points of $\mathcal{S}$ and  which are exact when restricted on $\mathcal{S}$.\par
It is well known that the  
simple Lie algebras  $A_n$, $B_n$, $C_n$ $D_n$ can be realized  as the matrix
Lie algebras   $\mathfrak{sl}(n+1,\mathbbm{C})$, $\mathfrak{so}(2n+1,\mathbbm{C})$,
$\mathfrak{sp}(2n,\mathbbm{C})$, $\mathfrak{so}(2n,\mathbbm{C})$. Correspondingly  there exist a loop representation of  $\widehat{\mathfrak{g}}$ on the infinite dimensional linear space $C^\infty(S^1,\mathbbm{C}^N)$ (where $N=n+1,2n+1,2n$ if $\mathfrak{g}$ is $A_n$, $B_n$, $C_n$ or $D_n$ resp.) \cite{CHPR}.
In this setting it is immediate to observe that  the one form $V(\lambda)$ is a solution of equation \rref{Caseq} if and only if as operator  in $\mbox{End}(C^\infty(S^1,\mathbbm{ C}^N))$ commutes with the linear differential operator $-\partial_x+S+\lambda A$ at any point $S\in \mathcal{S}$ \cite{CFMP3}. 
To find the elements $V(\lambda)$  commuting with $-\partial_x+S+\lambda A$ let us observe that  
$\Lambda=(B+\lambda A)$  can be viewed as an element of the tensor Lie algebra $\widetilde{\mathfrak{g}}=\mathfrak{g}\otimes \mathbbm{C}[\lambda,\lambda^{-1}]$ where  $\mathbbm{C} [\lambda,\lambda^{-1}]$ is the 
the commutative ring  of Laurent polynomials in the indeterminate $\lambda$ 
over $\mathbbm{C}$. In this algebra $\Lambda$ 
is a regular semisimple element  and therefore its  isotropic subalgebra  $\mathfrak{g}_{B+{\lambda} A}$  
is  a (Heisenberg) subalgebra $\mathfrak{H}$ of $\widetilde{\mathfrak{g}}$ 
spanned   in the case of the Kac--Moody Lie algebra of type $A^{(1)}_n$, $B^{(1)}_n$, $C^{(1)}_n$, by the matrices \cite{DS}:
\par
\begin{tabular}{llll}
 $\Lambda^{m}$&$\quad   m\in \mathbbm{Z}$ & $m\neq 0\mbox{mod}(n+1)$ &  if $\mathfrak{g}$ is of type $A_n$ \\
 $\Lambda^{2m+1}$&$\quad m\in \mathbbm{Z}$ & where for $m<0$,
$\Lambda^{2m+1}\stackrel{def}{=}\lambda^{-k}\Lambda^{2m+1+2nk}$ & if $\mathfrak{g}$ is of type $B_n$ \\
$\Lambda^{2m+1}$&$\quad m\in \mathbbm{Z}$ &  & if $\mathfrak{g}$ is of type $C_n$.\\
 \end{tabular}   
\par\noindent 
While in the more complicate case of the Lie algebra  $D^{(1)}_n$ the ``Heisenberg'' Lie algebra  $\mathfrak{H}$ is spanned by the matrices 
$$\Lambda^{2m+1} \quad m\in \mathbbm{Z} \quad \mbox{ where for $m<0$},\quad 
\Lambda^{2m+1}\stackrel{def}{=}\lambda^{-k}\Lambda^{2m+1+(2n-2)k}
$$
 together with the matrices 
 $$
\lambda^{m}F \quad m\in \mathbbm{Z},\quad F=\Phi+(-1)^n\Phi^T
\quad \Phi= e_{n,1}-  2e_{n+1,1}-2e_{n,2n} +4e_{n+1,2n},
$$
 here  $e_{ij}$ is the   matrix in $M(N,\mathbbm{C})$ with $1$ in the $ij$ position and zero otherwise and $X^T$ denotes the transpose of $X$.
From these facts Drinfeld and Sokolov proves indeed  the
\begin{prop}\label{DSZS} Let $\widetilde{G}$ be the Kac--Moody group $\widetilde{G}=C^\infty(S^1,G\otimes \mathbbm{C}[\lambda,\lambda^{-1}])$. Then for any operator of the form  $-\partial_x+S+\lambda A$ with $S\in \mathcal{S}$ there exists
a element $T$ in $\widetilde{G}$ such that:
\beq
\label{DSSZeq}
T(-\partial_x+S+\lambda A)T^{-1}=\partial_x+(B+\lambda A)+H, \quad H\in \mathfrak {H}.
\eeq
Therefore the set of the elements in $\widehat{\mathfrak{g}}\otimes \mathbbm{C}[\lambda,\lambda^{-1}]$ commuting with $-\partial_x+S+\lambda A$ is given (up the central charge) by
$T^{-1}{\mathfrak{H}}T$. 
\end{prop}
From Proposition \ref{DSZS} follows 
\begin{prop}\label{gencassol} Let $Z\in \mathfrak{H}$
 then: 
\begin{enumerate}
\item The element $V_Z=T^{-1}CT$ solves equation \rref{Caseq}.
\item Its hamiltonian on $\mathcal{S}$ is the function $H_Z=\langle K,Z\rangle$ where $K$ is defined by 
the relation 
\beq
\label{mommap}
K=T(S+\lambda A)T^{-1}+T_xT^{-1}.
\eeq
\item In particular:  if $Z$ is  $\Lambda^j$, $j=Nq+r$ with $q\in \mathbbm{Z}$ and $0<r<N$, then $V_{\Lambda^j}$   has the Laurent expansion
\begin{enumerate}
\item[]
\beq\label{VZ}
V_{\Lambda^j}= \lambda^q\sum_{p\geq -2}\frac{1}{\lambda^{p+1}}V^{Np+r}_1;  
\qquad  \mbox{ if\ \   $\widetilde{\mathfrak{g}}$ is of the type $A^{(1)}_n$, or $C^{(1)}_n$}
\eeq
 \item[] 
\beq\label{VZb} \hskip -1.95truecm 
V_{\Lambda^j}= \lambda^q\sum_{p\geq -2}\frac{1}{\lambda^{p+1}}V^{(N-1)p+q+r}_1; 
\qquad \mbox{if $\widetilde{\mathfrak{g}}$ if of type $B^{(1)}_n$ }
\eeq
\item[] 
 \beq\label{VZd} \hskip -1.78truecm 
V_{\Lambda^j}= \lambda^q\sum_{p\geq -2}\frac{1}{\lambda^{p+1}}V^{(N-2)p+2q+r}_1; 
\qquad \mbox{ if $\widetilde{\mathfrak{g}}$ if of type $D^{(1)}_n$ }
\eeq 
\end{enumerate}
where $V^{k}_1$ denote the coefficient of $\lambda^{-1}$ of  $V_{\Lambda^{k}}$.
 \item Finally if   $Z$ is  $\lambda^jF$, and therefore $\mathfrak{g}$ is of the type $D_n$ 
then   $V_{\lambda^{j}F}$  has the Laurent expansion 
$$
 V_{\lambda^{j}F}=\lambda^j\sum_{p\geq -2}\frac{1}{\lambda^{p+1}}V^{j+p-1}_1; 
$$
where $V^{k}_1$ denote the coefficient of $\lambda^{-1}$ of  $V_{\lambda^{k}F}$.
\end{enumerate}
 \end{prop}
{\bf Proof}\par\noindent
1. It follows immediately from the previous  proposition.\par\noindent
2. Using equation \rref{mommap} we can rewrite equation \rref{DSSZeq} in the form 
$T(-\partial_x+S+\lambda A)T^{-1}=-\partial_x+J$ showing the $K$ commutes with $C$ then:
$$
\frac{d}{dt}H_Z=\langle \dot{K},Z\rangle=\langle T\dot{S}T^{-1}+\left[\dot{T}T^{-1},K \right],Z\rangle
$$
but since $Z$ commutes with $K$ we have  
$$
\frac{d}{dt}H_Z=\langle T\dot{S}T^{-1},Z\rangle=\langle \dot{S},T^{-1}Z T\rangle=\langle \dot{S},V_Z\rangle.
$$
3.  First we observe that $T$ may be chosen of   the form \cite{DS} \cite{BdGHM}
$$
T=\mbox{exp}(\mathfrak{t})
\qquad\mbox{with}\ \ \ 
\mathfrak{t}=\sum_{k\geq 0}T_k\lambda^{-k}
$$
and therefore that $V_{\lambda^j}$ has the expansion
$$
 V_{\Lambda^j}=\lambda^{m(j)}V^j_{m(j)}+\sum_{k\geq -m(j)}V^j_k\lambda^{-k},
$$
where $m(j)=\left[\frac{j+n}{N}\right]$ if $\widetilde{\mathfrak{g}}$ is $A_n^{(1)}$ or $C_n^{(1)}$, $m(j)=\left[\frac{j}{N}\right]+1$  if $\widetilde{\mathfrak{g}}$ is $B_n^{(1)}$, $m(j)=\left[\frac{j}{N-1}\right]+1$
if $\widetilde{\mathfrak{g}}$ is $D_n^{(1)}$, and $\left[x\right]$ denotes the integer part of $x$.
Then for example equation \rref{VZd} follows from 
$$
\begin{array}{ll}
V^{j}_{p-q+1}&=\mbox{res}(\lambda^{p-q}V^j)=\mbox{res}(\lambda^{p-q}V^{Nq+r})=
\mbox{res}(\lambda^{p-q}T\Lambda^{Nq+r} T^{-1})\\
&=\mbox{res}(\lambda^{p-q}\lambda^qT\Lambda^{2q+r} T^{-1})  
=\mbox{res}(T\lambda^{p}\Lambda^{2q+r} T^{-1})=\mbox{res}(T \Lambda^{(N-2)p+2q+r}T^{-1})\\
&=\mbox{res}(V_{\Lambda^{(N-2)p+2q+r}}).
\end{array}
$$
While similar computations prove formulas \rref{VZ} and  \rref{VZd}.\par\noindent
4. It is almost trivial.
\endpf
 Using this proposition it is easy to give a bihamiltonian formulations of the hierarchies.
\begin{lem}\label{hierl}  The Drinfeld Sokolov hierarchies corresponding to the element $\Lambda^k$ ($k\in \mathcal{N}_{\widehat{\mathfrak{g}}}$) can be written in the bihamiltonian form:
 \beq
\label{bihfl}
\dot{S}_k=\left[A,V^k_1\right]=((V_{\Lambda^k})_+)_x+\left[(V_{\Lambda^k})_+,S+\lambda A\right]\qquad k\in \mathcal{N}_{\widehat{\mathfrak{g}}}.
\eeq
where $V^k_1$ is the residuum of $V_{\Lambda^k}$ and  $(V_{\Lambda^k})_+$ and $(V_{\Lambda^k})_-$ are respectively the projection on the regular and singular part of the Laurent
series $V_{\Lambda^k}$.\end{lem}
{\bf Proof} Using the expansion of $V_{\Lambda^j}$ in Proposition \ref{gencassol} it easy to see that any flow of the hierarchy may be written as
\beq
\label{jhyr}
\dot{S}_k=\left[A,V^k_1\right]\qquad k\in \mathcal{N}_{\widehat{\mathfrak{g}}}.
\eeq
Then since $V_{\Lambda^k}$ is a solution of \rref{Caseq} we have   that 
$$
((V_{\Lambda^k})_+)_x+\left[(V_{\Lambda^k})_+,S+\lambda A\right]=-((V_{\Lambda^k})_-)_x-\left[((V_{\Lambda^k})_-,S+\lambda A\right].
$$
 This latter equation   implies that
$$
\left[A,V^k_1\right]=((V_{\Lambda^k})_+)_x+\left[((V_{\Lambda^k})_+,S+\lambda A\right]\qquad k\in \mathcal{N}_{\widehat{\mathfrak{g}}}.
$$
\endpf
Observe that, in the present picture,  in the case of $D^{(1)}_n$ is missing the integrable hierarchy corresponding to the element $F$. Such hierarchy has been recently described and constructed by  Liu, Wu, and  Zhang in their  beautiful paper \cite{LWZ}.
\subsection{The Riccati equations}
We could now go ahead along this path of thoughts, by computing explicitely the matrices $V_{\Lambda^j}$ and projecting 
 them on the quotient space $\mathcal{N}$ to find the Drinfeld Sokolov hierarchies.
However we shall see that there exists a 
different and maybe easier way to achieve  the same result.\par
To begin with, let us find an eigenvector and an eigenvalue of the the linear differential operators differential operators $-\partial_x+S+\lambda A$.
\par\noindent  
 Let $\{e_j\}_{j=0,\dots,N-1}$ with
$$
e_0=(1,0,\dots,0)^T\quad e_j=(\underbrace{0,\dots,0}_{j-1},1,0,\dots,0)^T \qquad j=0,\dots N-1
$$ 
 the canonical basis of $\mathbbm{C}^N$ , 
 $(\cdot,\cdot)$  be the   pairing of $\mathbbm{C}^N$ defined by the relation $(e_i,e_j)=\delta_{ij}$ where  $\delta_{ij}$ is the usual Kronecker delta. 
Let further set $v^{(0)}=(1,0,\dots,0)^T\in C^\infty(S^1,\mathbbm{C}^N)$ and define recursively 
\beq
\label{recrelabc}
v^{(j+1)}(S)=\partial_xv^{(j)}(S)+(S+\lambda A)^Tv^{(j)}(S)\qquad (v^{(0)}(S)=v^{(0)}),
\eeq 
then it holds 
 \begin{theorem}
\label{eigpro}  
\begin{enumerate}\item  Let $\mathfrak{g}$ be 
 the  simple Lie algebra   $A_n$, $B_n$ or $C_n$ then 
\begin{enumerate}
\item[a)]
the subset $\{v^{(j)}\}_{j=0,\dots,N-1}$
is for any $S$ in $\mathcal{S}$ a basis for ${\Bbb C}^N$. 
We may therefore 
develop  the  (for any fixed $S$) first dependent vector, namely $v^{(N)}(S)$,  obtaining  the relation
\beq
\label{careq} 
v^{(N)}(S)=\sum_{k=0}^{N-1} c_k(S)v^{(k)}(S)
\eeq
 called the ``characteristic equation'' of the operator   $-\partial_x+S+\lambda A$.
\item[b)]  Let $\psi$ be   the element of $C^\infty(S^1,{\Bbb C}^N)$ defined by the relations  
$\langle v^{(0)},\psi\rangle=1$, $\langle v^{(1)},\psi\rangle=h$ and  $\langle v^{(k)},\psi\rangle=h^{(k)}$
$k=2,\dots,N-1$,  where the function $h^{(k)}$ are defined by the recurrence: $h^{(1)}=h$, $h^{(k+1)}=h^{(k)}_x+h^{(k)}h$. Then if 
 $h$ satisfies  the ``Riccati--type'' equation
\beq
\label{Riceqn}
h^{(N)}=\sum_{k=0}^{N-1}c_k(x,\lambda)h^{(k)}
\eeq
whose coefficients $c_k(x,\lambda)$ are those of equation \rref{careq}, $\psi$ is an eigenvector of $-\partial_x+S+\lambda A$ with eigenvalue $h(z)$. 
 \end{enumerate}
\item  Let $\mathfrak{g}$ be 
the simple Lie algebra  $D_n$ then
\begin{enumerate} 
\item[a)]  the subset $\{v^{(j)}\}_{j=0,\dots,2n-2}$ together with the vector 
$$
w^{(n)}=(\underbrace{0,\dots,0}_{n}1,0,\dots,0)^T
$$
is for any $S$ in $\mathcal{S}$ a basis for ${\Bbb C}^{2n}$, developing the first 
dependent vector $v^{(2n-1)}$ with respect to this basis 
we obtain the relation 
\beq
 \label{careqd} 
v^{(2n-1)}(S)=\sum_{k=0}^{2n-2}c_k(S)v^{(k)}(S)+c_w(S)w^{(n)}.
\eeq
\item[b)] There exist functions $d_i\in C^\infty(\mathcal{S},\mathbbm{C})$, $0\leq i\leq n-1$ such that if 
$\psi\in C^\infty(S^1,{\Bbb C}^{2n})$ is defined by the relations  
$\left(v^{(0)},\psi\right)=1$, $\left( v^{(1)},\psi\right)=h$,  
$\left( v^{(k)},\psi\right)=h^{(k)}$
$k=2,\dots,N-1$,  and $\left( w^{(n)},\psi\right)=\sum_{i=0}^{n-1} d_ih^{(i)}-c_wh^{(-1)}_{d_0}$ 
where the functions $h^{(k)}$ are defined as above, and $h^{(-1)}_{d_0}$ is defined by the relation $(\partial_x+h)(h^{(-1)}_{d_0})=d_0$, and   
 $h$ satisfies  the ``Riccati--type'' equation
\beq
\label{Riceqnd}
h^{(2n-1)}=\sum_{k=0}^{n-1}(c_k(x,\lambda)-c_wd_k)h^{(k)}+
\sum_{k=n}^{2n-2}c_k(x,\lambda)h^{(k)}
-c_wh^{(-1)}_{d_0}
\eeq
where  the coefficients 
$c_k(x,\lambda)$ are those of equation \rref{careq}, then $\psi$ is an eigenvector of $-\partial_x+S+\lambda A$ with eigenvalue $h(z)$.
\end{enumerate}
\end{enumerate}
\end{theorem}  
{\bf Proof}\par\noindent 
1.a)  Using the  ``canonical'' basis of $\mathbbm{C}^N$:
$\{e_j\}_{j=0,\dots, N-1}$ and the 
form of the matrices $S+\lambda A$ with to respect to this basis 
 we have easily  
$$
v^{(j)}=e_j+\sum_{k<j}c_k(x)e_k\qquad j=0,\dots N-1,
$$
which proves the linear independence  of the first $N$ elements $v^{(j)}$, $j=0,\dots N-1$.\par\noindent
1.b)
Pairing the relation $-\partial_x+(S+\lambda A)\psi=h\psi$ with the vectors $v^{(j)}(S)$ $j=0,\dots , N-1$ we obtain:
$$
-\left(v^{(j)},\partial_x \psi\right)+\left((v^{(j)},(S+\lambda A)\psi\right)=h\left(v^{(j)},\psi\right)\qquad 
j=0,\dots, N-1
$$
which can be written as
$$
\begin{array}{ll}
&-\partial_x\left(v^{(j)},\psi\right)+\left(\partial_xv^{(j)},\psi\right)
+\left((S+\lambda A)^Tv^{(j)},\psi\right)\\
&=-\partial_x\left(v^{(j)},\psi\right)+\left((v^{(j+1)},\psi\right)
=h\left(v^{(j)},\psi\right)\qquad 
j=0,\dots, N-1
\end{array}
$$
These latter equations are for $j=0,\dots, N-2$ satisfied if 
$$
h^{j+1}=\left((v^{(j+1)},\psi\right)=\partial_x\left(v^{(j)},\psi\right)+h\left(v^{(j)},\psi\right)
=h^{(j)}_x+hh^{(j)}
$$ 
as required in the Hypothesis, while for $j=N-1$ using \rref{careq} we have 
$$
\begin{array}{ll}
h^{(N)}&=\left((v^{(N-1)},\psi\right)_x+h\left((v^{(N-1)},\psi\right)=
\left((v^{(N)},\psi\right)\\
&=\left(\sum_{k<N}c_k(S)v^{(k)}(S),\psi\right)=\sum_{k=0}^{N-1}c_k(x,\lambda)h^{(k)}.
\end{array}
$$
2. a) Using again the canonical basis $\{e_j\}_{j=0,\dots, 2n-1}$ and the 
form of the matrices $S+\lambda A$ with to respect to this basis we have
$$\begin{array}{lll}
v^{(j)}&=e_j+\sum_{k<j}c_k(x)e_k&\qquad j=0,\dots n-2,\\
v^{(n-1)}&=\frac{1}{2}e_{n}+e_{n-1}+\sum_{k\leq n-2}c_k(x)e_k&\\
v^{(j)}&=e_j+\sum_{k<j}c_k(x)e_k&\qquad j=n,\dots 2n-2\\
\end{array}
$$
and therefore together with the element $w^{(n)}$ form a basis for $\mathbbm{C}^{2n}$ at any point $S$ of $\mathcal{S}$.\par\noindent
2. b) Reasoning like in the point 1. b) we have that $\psi$ is an eigenvector 
of $-\partial_x+S+\lambda A$ with eigenvalue $H$ if and only if $h$ satisfies the relation
$$
h^{(2n-1)}=\sum_{k=0}^{2n-2}c_kh^{(k)}+c_w\left(w^{(n)},\psi\right).
$$
Now since $\mathcal{S}-B\subset p_\theta(x)H_\theta +C^\infty(S^1,\mathfrak{n}^-)$ 
and it easily checked that $H_\theta w^{(n)}=0$ and $Bw^{(n)}=e_{n+1}$ it yields
$$
\begin{array}{ll}
(\partial_x+h)\left(w^{(n)},\psi\right)&= \left(w^{(n)},\psi\right)_x+\left(w^{(n)},h\psi\right)=\left(w^{(n)},\psi\right)_x+\left(w^{(n)},(-\partial_x+S+\lambda A)\psi\right)\\&=\left((-\partial_x+(S+\lambda A)^Tw^{(n)},\psi\right)=\left(v^{(n)}-\sum_{i=0}^{n-2}f_i(x)v^{(i)},\psi\right)\\
&=h^{(n)}-\sum_{i=1}^{n-2}f_i(x)h^{(i)}(x)-f_0(x).
\end{array}
$$
Therefore  $\left(w^{(n)},\psi\right)$ must have the form 
$$
\left(w^{(n)},\psi\right)=h^{(n-1)}-\sum_{i=1}^{n-2}d_i(x)h^{(i)}(x)-h^{(-1)}_{d_0}.
$$
\endpf
To explicitely compute the coefficients of the Riccati equations we shall use 
\begin{prop}\label{recprop} The vectors $v^{(j)}(S)$ $j\geq 0$ are covariant 
i.e.,  $v^{(j)}(S')=(J^T)^{-1}v^{(j)}(S)$ whenever  $S'=JSJ^{-1}-J_xJ^{-1}$ with $J\in G_{AB}$. 
\end{prop} 
{\bf Proof}
The first statement follows by induction over $j$. For $j=0$ we have indeed\\ $(J^T)^{-1}v^{(0)}=v^{(0)}$ from the definition of the group $G_{AB}$. While if the statement is true for $i$ for $i+1$ we have
$$
\begin{array}{ll}
&v^{(i+1)}(S')=\partial_xv^{(i)}(S')+(S'+\lambda A)^Tv^{(i)}(S')=
\partial_x((J^T)^{-1}v^{(j)}(S))\\&+(J^T)^{-1}(S^T+\lambda A^T)J^T(J^T)^{-1}v^{(i)}(S)-(J^T)^{-1}J_x^T(J^T)^{-1}v^{(i)}(S)=
(J^T)^{-1}J_x^T(J^T)^{-1}v^{(i)}(S)\\&+(J^T)^{-1}(S^T+\lambda A^T)v^{(i)}(S)-(J^T)^{-1}J_x(J^T)^{-1}Tv^{(i)}(S)=(J^T)^{-1}
(S+\lambda A)^Tv^{(i)}(S).\hskip 1truecm \square
\end{array}  
$$
Thus the coefficients of the Riccati equations are 
 invariant under the gauge action of the group $G_{AB}$, 
and can be  computed   using  the transversal manifold  $\mathcal{Q}$ \ref{smtra}, which in our setting  can be characterized as follows \cite{DS} \cite{CP} \cite{DLZ}.\par  
Let $\widehat{\mathfrak{g}}$ be a affine Lie algebra of the type specified above and let  $\mathfrak{g}$ be the corresponding  simple Lie  algebras then  the map  \cite{Ko}
$$
ad_B : \mathfrak{n}^-\to \mathfrak{b}=\mathfrak{n}^-\oplus \mathfrak{h}
$$
is injective. We fix a subspace $\mathfrak{q}$  of $\mathfrak{b}$  such that
\beq
\label{V}
\mathfrak{b} = \mathfrak{q} \oplus \left[B, \mathfrak{n}\right],
\eeq
so $\mbox{dim}(\mathfrak{q}) = \mbox{dim}(\mathfrak{b}) -\mbox{dim}(\mathfrak{n}) = n=\mbox{rank}\mathfrak{g}$. Then the manifold:
$$
\mathcal{Q}=C^\infty(S^1,\mathfrak{q})
$$
is the  transversal manifold defined in section 2.2.
To determine it explicitely let consider the decomposition of $\mathfrak{g}$  w.r.t. the principal gradation \cite{DLZ} \cite{CP}
$$
\mathfrak{g}=\bigoplus_{1-h\leq j\leq h-1}\mathfrak{g}^{j}\qquad  \mathfrak{g}^j=\left\{\begin{array}{lll} \mathfrak{h}& & \mbox{if $j=0$}\\
 \bigoplus_{\rm{ht}(\alpha)=j}\mathfrak{g}_\alpha & &\mbox{if $j\neq 0$}
\end{array}\right.
$$
where $\rm{ht}$ is the the height function of roots: $\rm{ht}(\alpha)=\sum_{\alpha_i\in \Pi} n_i\alpha_i$ and $h$ is the Coxeter number of $\mathfrak{g}$.
Then we specify the choice of the complement $\mathfrak{q}$ of the subspace $\left[I, \mathfrak{n}\right]$ of $\mathfrak{b}$  so that \cite{CP} \cite{DLZ}:
$$
\mathfrak{q}=\bigoplus_{j=o}^{h-1}\mathfrak{q}_j
$$
where the subspaces $\mathfrak{q}_j$ satisfy
$$
\mathfrak{q}_j\subset \mathfrak{b}_j=\mathfrak{b}\cap\mathfrak{g}^j,\qquad 
\mathfrak{b}_j= \mathfrak{q}_j\oplus \left[B, \mathfrak{b}_{j+1}\right].
$$
Note that $\mathfrak{q}_j$ is not a null space if and only if $j$ is one of the exponents
$$
1 = m_1\leq m_2 \leq\cdots\leq  m_n = h-1
$$
of the simple Lie algebra $\mathfrak{g}$. For all simple Lie algebras except the ones of $D_n$ type with even $n$
the exponents have multiplicity one, i.e. $\mbox{dim}(\mathfrak{q}_{m_i})
= 1$ and the exponents are distinct. For the $D_n$
(with even $n$) case, the exponents $m_i$ for $i \neq  \frac{n}{2},
, \frac{n}{2}+1$ have multiplicity one, $m_\frac{n}{2}
= m_{\frac{n}{2}+1} = n-1$ and $\mbox{dim}(\mathfrak{q}_{n-1}) = 2$.
Using these facts it is not difficult to determine a transversal manifold.  Denoting with $e_{ij}$, $i.j=0,\dots N-1$ the $N\times N$ matrix with $1$ in the entry $(i,j)$ and zero elsewhere we have indeed:\par\noindent
\begin{tabular}{c|l} 
Kac--Moody algebra   & Tranversal manifold $\mathcal{Q}=C^\infty(S^1,\mathfrak{q})$\\
\hline
$A_n^{(1)}$ & $\mathfrak{q}=B+\sum_{i=0}^{n-2}u_i(x)e_{n-1,i}$\\
 $B_n^{(1)}$  & $\mathfrak{q}=B+\sum_{i=0}^{n-1}u_i(x)(e_{2n-1-i,i}+e_{2n-i,i+1})$\\
 $C_n^{(1)}$ & $\mathfrak{q}=B+\sum_{i=0}^{n-1}u_i(x)e_{i,2n-1-i}$\\
 $D_{2m}^{(1)}$ & $\mathfrak{q}=B+\sum_{i=0}^{m-1}u_i(x)(e_{4m-2-i,i}+e_{4m-1-i,i+1})$ \\
&  $+u_m(e_{3m-2,m-2}+e_{m-1,3m+1})+\sum_{i=m}^{2m-2}u_{i+1}(x)(e_{4m-2-i,i}+e_{4m-1-i,i+1})$
 \\
$ \ \ \ D_{2m+1}^{(1)}$ & $\mathfrak{q}=B+\sum_{i=0}^{m-1}u_i(x)(e_{4m-i,i}+e_{4m+1-i,i+1})$\\
&   $+u_m(e_{3m,m-1}e_{m-1,3m+1})+ \sum_{i=m}^{2m-2}u_{i+1}(x)(e_{4m-i,i}+e_{4m+1-i,i+1})$
\end{tabular}\vskip 1truecm \noindent
For example we have that the transversal manifold for the Lie algebra $C_2^{(1)}$ (resp.   $B_3^{(1)}$) is
$$
\mathfrak{q}_{C^{(1)}_2}=\left(\begin{array}{cccc} 0 & 1 & 0 & 0\\0 & 0 & 1 & 0\\ 0 & u_1 & 0 &1 \\ 
                    u_0 & 0 & 0& 0\end{array}\right)
\qquad\mathfrak{q}_{B^{(1)}_3}=\left(\begin{array}{ccccccc} 0 & 1 & 0 & 0 & 0 & 0 & 0\\
0 & 0 & 1 & 0& 0 & 0& 0 \\ 0 & 0 & 0 &1 & 0 & 0 & 0 \\ 
                   0 & 0 &  u_2 & 0 & 1& 0 & 0\\
0 & u_1& 0 & u_2 & 0 & 1 & 0\\
u_0& 0 & u_1 & 0 & 0 & 0 & 1\\
0 & u_0& 0 & 0 & 0& 0& 0\end{array}\right)
$$
while  the transversal manifold for the Lie algebra $D_3^{(1)}$ (resp.   $D_4^{(1)}$) is
$$
\mathfrak{q}_{D_3^{(1)}}=\left(\begin{array}{cccccc} 0 & 1 & 0 & 0& 0 & 0\\0 & 0 & 1 & \frac{1}{2}& 0 & 0\\0 & 0 & 0 & 0 & \frac{1}{2} & 0\\
 u_1 & u_2 &0 & 0&1 &0 \\ 
u_0 & 0 & u_2& 0&0 & 1\\
0 &u_0&-u_1  & 0& 0&0 
\end{array}\right)
\qquad\mathfrak{q}_{D_4^{(1)}}=\left(\begin{array}{cccccccc} 0 & 1 & 0 & 0 & 0 & 0 & 0 &0\\
0 & 0 & 1 & 0& 0 & 0& 0 &0\\ 0 & 0 & 0 &1 & \frac{1}{2} & 0 & 0 &0\\ 
 0 & 0 & 0 &0 & 0  & \frac{1}{2} & 0 &0\\               
   u_2 & 0 &  u_3 & 0 & 0& 1 & 0 &0\\
0 & u_1& 0 & u_3 & 0 & 0 & 1 & 0\\
u_0& 0 & u_1 & 0 & 0 & 0 & 0 &1\\
0 & u_0& 0 & u_2 & 0& 0& 0 & 0\end{array}\right).
$$
Using the explicit form of the transversal manifold it is easy to check that $c_0=\lambda+g_0(Q)$  and all the other coefficients $c_k$ are independent  if $\widehat{\mathfrak{g}}$ is of the type $A^{(1)}_n$ 
or  $C^{(1)}_n$ and that $c_1=\lambda +g_1(Q)$ and all the other coefficients  are independent of $\lambda$ if $\widehat{\mathfrak{g}}$ is of the type $B^{(1)}_n$ or $D^{(1)}_n$. Moreover if $m$  is the index of the highest Fa\`a di Bruno polynomial appearing in the Riccati equation then $c_{m-1}=0$. For example the first Riccati equations are
$$\begin{array}{lll}
h^{(2)}=&\lambda+u_0&\qquad \mbox{ if  $\widehat{\mathfrak{g}}$  is $A_1^{(1)}$}\\
h^{(4)}=&u_1h^{(2)}+ u_{1x}h+\lambda+u_0&\qquad \mbox{ if  $\widehat{\mathfrak{g}}$  is $C_1^{(2)}$}\\
h^{(7)}=&2u_2h^{(5)}+5u_{2x}h^{(4)}+(2u_1+4u_{2xx})h^{(3)}+(3u_{1x}+u_{2xxx})h^{(2)} & \\&+
(\lambda+2u_0+u_{1xx})h+u_{0x}&\qquad \mbox{ if  $\widehat{\mathfrak{g}}$  is $B_1^{(3)}$}\\
h^{(5)}=&u_2h^{(3)}+\frac{3}{2}u_{2x}h^{(2)}+(\lambda+2u_0-u_{1x}+\frac{1}{2}u_{2xx}-\frac{1}{4}u_2^2)h^{(1)} &
\\&+(u_{0x}-\frac{1}{2}u_{1xx}-\frac{1}{4}u_2u_{2x})+\frac{1}{4}(u_{2x}+u_1)h^{(-1)}_{u_{2x}+u_1} &\qquad \mbox{ if  $\widehat{\mathfrak{g}}$  is $D_1^{(3)}$}\\
h^{(7)}=&u_3h^{(5)}+\frac{5}{4}u_{3x}h^{(4)}+(2u_{3xx}+2u_1+u_2-\frac{1}{4}u_3^2)h^{(3)} &\\&+
(3u_{1x}+\frac{3}{2}u_{2x}-\frac{3}{4}u_3u_{3x}+\frac{1}{2}u_{3xxx})h^{(2)} &\\&+
(\lambda+2u_0+u_{1xx}+\frac{3}{2}u_{2xx}-\frac{1}{2}u_{2}u_3-\frac{1}{4}u_{3xx}u_3)h^{(1)} &\\&+(u_{0x}-\frac{1}{4}u_{3x}u_2-\frac{1}{4}u_{2x}u_{3})
+\frac{1}{2}u_{2xxx}+\frac{1}{4}u_{3xx}u_{3x})&\\&
-\frac{1}{4}(u_{3xx}+u_2)h^{(-1)}_{u_{3xx}+u_2} &\qquad \mbox{ if  $\widehat{\mathfrak{g}}$  is $D_1^{(5)}$}\\
 \end{array}
$$
 \begin{prop}\label{hsolve}  Any Riccati equation \rref{Riceqn} or \rref{Riceqnd}
admits a solution of the form  $h(z)=z+\sum_{i< 0}h_iz^{-i}$ where $z^{n+1}=\lambda$ if $\widehat{\mathfrak{g}}=A_n^{(1)}$, $z^{2n}=\lambda$ if \
$\widehat{\mathfrak{g}}=B_n^{(1)}$,  or $C_n^{(1)}$ and finally $z^{2n-2}=\lambda$ if \ $\widehat{\mathfrak{g}}=D_n^{(1)}$
whose  coefficients $h_k$ are obtained iteratively in a pure  algebraic way.
\end{prop}
 {\bf Proof} 
It is immediate to show by induction that if $h(x)$ is a formal Laurent series of the form $h(z)=z+\sum_{i> 0}h_iz^{-i}$   then $h^{(k)}$ has the form
$$
h^{(k)}=z^{k}+\sum_{j\geq -k+1}h^k_jz^{-j}\qquad \mbox{with}\quad  
h^k_j=kh_{j-k+1} +q^k_j
$$
where $q^j_k$ is a differential polynomial in the coefficients $h_i$ $i=1,\dots,
 j-1$. Therefore by substituting these expressions in equation \rref{Riceqn} and developing it in powers of $z$  we obtain that the equation corresponding to the powers $N$ and $N-1$ are automatically satisfied while that corresponding to the power $i$, $i\leq N-2$ is of the type
$$
Nh_{N-i-1}=\mbox{differential polynomial in $h_1,\dots,h_{N_i-2}$ and  $u_0,\dots,u_{n-1}$}.
$$
The same argument works in  the case of the equation  \rref{Riceqnd} once  we have shown that condition $(\partial_x+h)h^{(-1)}_g(u)=g(u)$ determines completely $h^{(-1)}_g(u)$
as Laurent series $h^{(-1)}_g(u)=k_1z^{-1}+\sum_{j\geq 1}k_iz^{-i}$, whose coefficients $k_i$ are differential polynomials in the function $g(u)$ and in the coefficients $h_j$ with $j<i$, which, however, still follows immediately by induction.   
\endpf
We can write the Riccati equation in a more compact form using the following
\begin{prop}\label{skew} The  Riccati equations   \rref{Riceqn} and  \rref{Riceqnd} can be written as
$$ 
\begin{array}{lll}
h^{(n+1)}&=\lambda+\sum_{k=0}^{n-1}u_kh^{(k)}  &\qquad \mbox{ if \ \ $\widehat{\mathfrak{g}}=A_n^{(1)}$}\\
h^{(2n+1)}&=\lambda h^{(1)}+\sum_{k=0}^{n-1}u_kh^{(2k+1)}+\sum_{k=0}^{n-1}(\partial_x+h)^{(2k+1)}(u_k) &\qquad \mbox{ if \ \ $\widehat{\mathfrak{g}}=B_n^{(1)}$}\\
h^{(2n)}&=\lambda+\sum_{k=0}^{n-1}u_kh^{(2k)}+\sum_{k=0}^{n-1}(\partial_x+h)^{(2k)}(u_k)  &\qquad \mbox{ if \ \ $\widehat{\mathfrak{g}}=C_n^{(1)}$}\\
h^{(2n-1)}&=\lambda h^{(1)}+\sum_{k=1}^{n-1}u_kh^{(2k-1)}+\sum_{k=0}^{n-1}(\partial_x+h)^{(2k-1)})u_k) + u_0h^{(-1)}_{u_0}&\qquad \mbox{ if \ \ $\widehat{\mathfrak{g}}=D_n^{(1)}$}
\end{array}
$$
\end{prop}
{\bf Proof} The form of the Riccati equation corresponding to the Lie algebra $A_n^{(1)}$ follows immediately from the very from of the transversal manifold $\mathcal{Q}$ in this case \cite{CFMP3}. \par 
In the other cases the operator $-\partial_x+(S+\lambda A)$ is skew symmetric with respect the bilinear form \cite{DS} 
$$
\begin{array}{ll}
\langle \cdot,\cdot\rangle_\Omega:C^\infty(S^1,\mathbbm{C}^N)&\to \mathbbm{C}\\
(v,w)\mapsto\langle v,w\rangle_\Omega&= \int_{S^1}(v,\Omega w)dx
\end{array}
$$
where $\Omega=\mbox{diag}(1,-1,\dots,-1, 1)$ if $\widetilde{\mathfrak{g}}=B^{(1)}_n$,  $\Omega=\mbox{diag}(1,-1,\dots,1, -1)$ if 
$\widetilde{\mathfrak{g}}=C^{(1)}_n$, and finally $\Omega=\mbox{diag}((1, -1\dots,(-1)^{n-1},(-1)^{n-1},(-1)^{n},\cdots, 1)$ if $\widetilde{\mathfrak{g}}=D^{(1)}_n$. This in turn implies using the proof of Theorem \ref{eigpro} that if h(x) satisfies the relation \rref{Riceqn} (resp. \rref{Riceqnd})  it must also satisfies the relation $(-1)^{N-1}h^{N-1}=\sum_{k=0}^{N-1}(-1)^k(\partial_x+h)^k(c_k(u))$ (resp. 
$(-)^{N-1}h^{N-1}=\sum_{k=0}^{N-1}(-1)^k(\partial_x+h)^k(c_k(u))-d_0h^{(-1)}_{c_w}$).
Therefore with respect to the adjoint operation $(\partial_x+h)^*=-(\partial_x+h)$, $f(u)^*=f(u)$ the relation  (\ref{Riceqnd}  is skew symmetric while \rref{Riceqn}  is  skew symmetric  if $\mathfrak{g}$ is of type $B$  and  symmetric if $\mathfrak{g}$ is of type $C$). \endpf
\subsection{The construction of the hierarchies}
 Actually the proof of Theorem \ref{eigpro} and Proposition \ref{hsolve} 
shows  that 
$h_p=h(\omega^pz)$ and $\psi_p=\psi(\omega^pz)$  are respectively eigenvalues and eigenvectors of $-\partial_x+S+\lambda A$, with 
$\omega=\exp{\left(\frac{2\pi i}{N'}\right)}$  and $p=0,1,\dots,N'$, where $N'=n+1$ if $\widehat{\mathfrak{g}}=A_n^{(1)}$, $N'=2n$ if 
$\widehat{\mathfrak{g}}=B_n^{(1)}$, or $\widehat{\mathfrak{g}}=C_n^{(1)}$ and finally $N'=2n-2$ if $\widehat{\mathfrak{g}}=D_n^{(1)}$
 Moreover since the eigenvalues $h(\omega^pz)$ are distinct we have that  the corresponding eigenvectors are linear independent.  Further equation \rref{DSSZeq} shows that the operator $-\partial_x+(S+\lambda A)$ has together with the non trivial eigenvectors $\psi(\omega^pz)$ an eigenvector $\chi_0$ (resp. two eigenvectors $\chi_1$, $\chi_2$) with eigenvalue zero when $\mathfrak{g}$ is of type $B$ (resp. of type $D$). Therefore at any point of $S\in \mathcal{S}$ the eigenvectors $\psi(\omega^pz)$ $p=0,\dots N-1$ 
if $\widehat{\mathfrak{g}}=A_n^{(1)}$ or $\widehat{\mathfrak{g}}=C_n^{(1)}$,
$\psi(\omega^pz)$, $p=0,\dots N-1$, $\chi_0$ if $\widehat{\mathfrak{g}}=B_n^{(1)}$, and finally
 $\psi(\omega^pz)$, $p=0,\dots N-1$, $\chi_1$, $\chi_2$ if $\widehat{\mathfrak{g}}=D_n^{(1)}$, form a basis of $\mathbbm{C}^N$.
This fact allows us to prove 
\begin{theorem}\label{hev} The Drinfeld--Sokolov hierarchies  corresponding to the Casimirs of type $V_{\Lambda^j}$ can be written in the form
\beq
\label{heveq}
\partial_{t_j}h=\partial_xH^{(j)}\qquad j\in \mathcal{N}_{\widehat{\mathfrak{g}}}
\eeq
where the Laurent series $H^{(j)}$  are 
given by $H^{(j)}=\langle v^{(0)},(V_{\Lambda^j})_+\psi_0\rangle$.
\end{theorem}
{\bf Proof} 
We shall present here the proof in the case of the affine Lie algebra $D_n^{(1)}$ being the other cases with the obvious changes similar.\par
The Drinfeld--Sokolov hierarchies in this case can be written as \ref{hierl}
$$
\left[-\partial_x+(S=\lambda A),-\partial_t+(V_{\Lambda^{2j+1}})_+\right]=0\qquad j\in \mathbbm{N}
$$
which together with $-\psi_{0x}+(S+\lambda A)\psi_0=h_0\psi_0$
gives 
\beq
\label{phipsi}
(-\partial_x+S+\lambda A)\phi=h\phi-h_{t_{2j+1}}\psi_0
\eeq
where 
\beq
\label{phi}
\phi=(-\partial_{t_{2j+1}}+V_{\Lambda^{2j+1}})\psi_0.
\eeq
Decomposing now $\phi$ with respect to the basis $\chi_1,\chi_2,\psi_0,\dots,\psi_{2n-2}$: $\phi=a_1\chi_1+a_2\chi_2\chi+\sum_{l=0}^{2n-1} c_l\psi_l$,
 equation \rref{phipsi} becomes 
$$
-a_{1x}\chi_1 -a_{2x}\chi_2+\sum_{l=0}^{2n-1}(-c_{lx}+c_lh_l)\psi_l=ha_1\chi_1+ha_2\chi_2+\sum_{l=0}^{2n-1} hc_l\psi_l-h_{t_{2j+1}}\psi_0
$$
which implies  therefore $-a_{ix}=ha_i$ $i=1,2$ and $-c_{lx}-c_lh_l=c_lh$  ($h_0=h$) for $l=1,\dots,2n-2$. Then the Laurent series  $c_l$ $l=0,\dots 2n-1$ must fulfill the conditions $c_{lx}=c_l(h_l-h)$, and since  the Laurent series $h_l-h$ have degree $1$ in $z$ these imply $c_l=0$ if $l\neq 0$. Hence $\phi=c_0\psi$ which taking into account \rref{phi}, the definition of $H^{(j)}$
and the normalization of $\psi_0$ implies $c_0=H^{(j)}$. Therefore $\phi=H^{(j)}\psi_0$ which plugged in \rref{phipsi}
yields  $h_{t_j}=\partial_x H^{(j)}$. $\square$\par
We can now show the Laurent series $H^{(j)}$ can be computed directly from their eigenvalue $h(z)$ avoiding the construction of the matrix $(V_{\Lambda^j})_+$ or the use of the Poisson pencil.
 \begin{theorem}\label{Hcon} 
 The Laurent series  of $H^{(j)}$ is given by
\beq
\label{asbeHeq}
H^{(j)}=z^j+\sum_{l\geq 1}H^j_lz^{-l}.
\eeq
and has a Fa\`a di Bruno expansion 
\beq
\label{faexp}
H^{(j)}=h^{(j)}+\sum_{i=0}^{j-1} s^j_i(x)h^{(i)}
\eeq  
where the coefficients $s^j_i(x)$ are independent of $z$. Therefore if we define\\ 
$H_+=\mbox{span}\langle h^{(j)}\vert\ j\geq 0\rangle$
the functions $H^{(j)}$ are the projections
$$
H^{(j)}=\pi_+(z^j)
$$
of $z^j$ onto  $H_+$  along splitting of the space  of Laurent series in the direct sum of $H_+$ and of the subspace of strictly negative Laurent series.
\end{theorem}
{\bf Proof} From  formula (\ref{DSSZeq}) and Theorem \ref{eigpro} follows  that 
$T\psi_0$ is an eigenvector of $\Lambda$ with eigenvalue $z$ and hence $V_{\Lambda^j}\psi_0=z\psi_0$. Then from the definition of $H^{(j)}$ we have 
$H^{(j)}=\left( v^{(0)},(V_{\Lambda^j})_+\psi_0\right)=\left( v^{(0)},V_{\Lambda^j}\psi_0\right)-
\left( v^{(0)},(V_{\Lambda^j})_-\psi_0\right) =z^j-\left( v^{(0)},(V_{\Lambda^j})_-\psi_0\right)$. Now from the definition of the $h^{(k)}$ and the Riccati equations follows  that \\
$\psi_0=\left(1,z+O(1),z^2+O(z), \dots, z^{N-1}+O(z^{N-2})^T\right)$ and therefore   we have that
$$
-\left( v^{(0)},(V_{\Lambda^j})_-\psi_0\right)=\frac{H_1^j}{z}+\dots
$$
because in the case of the Lie algebras of type $A$ or $C$,
 $\Lambda=z^N$; while in the case of the Lie algebras of type   $B$ or $D$ we have 
$\left( v^{(0)},(V_{\Lambda^j})_-(\psi_0)_{N-1}\right)=0$, because $((V_{\Lambda^j})_-)_{0,N-1}=0$.\par  
Now if  $\mathfrak{g}$ is of either $A_n$ or $C_n$:
\beq\label{Hac}
\left( v^{(0)},(V_{\Lambda^j})_+\psi_0\right)=\left( (V_{\Lambda^j})_+^Tv^{(0)},\psi_0\right)=
\left(\sum_{i=0}^{N-1}r^j_iv^{(i)},\psi_0\right)=\sum_{i=0}^{N-1}r^j_ih^{(i)}
\eeq
where the coefficients $r^j_i$ are polynomials in $\lambda$. In this case using the  equation (\ref{Riceqn}) we get 
\beq\label{lambeqac}
\lambda=h^{(N)}-\sum_{k=0}^{N-2}c_k(x)h^{(k)}
\eeq
where the coefficients $c_k$ are independent of $\lambda$ and therefore that $\lambda\in H_+$. Moreover acting iteratively  on (\ref{lambeqac}) with $(\partial_x+h)$ we get 
$$
\lambda h^{(j)}=h^{(N+j)}-\sum_{k=0}^{N-j-1}c^j_k(x)h^{(k)}
$$
which shows $\lambda h^{(j)}\in H_+$ and by induction that  
$\lambda^ih^{(j)}\in H_+$, $i,j\in \mathbbm{N}$.  Therefore form equation (\ref{Hac}) it easily seen $h^{(j)}\in H_+$, $j\in \mathbbm{N}$.\par
Unfortunately in the case of  Lie algebras of type $B$ and $D$ we can not use the above argument, but  we need to introduce the map $\phi$ from the space of Laurent series to the space of pseudo differential operator defined in \cite{CFMP4}, which maps the Fa\`a di Bruno Polynomial $h^{(k)}$ to the operator $\partial_x^k$   together with the results obtained in such paper.  In particular  we have that $P_{2k+1}=\phi(H^{(2k+1)})$ is the unique pseudo differential operator  such that $P_{2k+1}=(L^{\frac{2k+1}{2n}})_+$ if $\widetilde{\mathfrak{g}}$ is $B_n^{(1)}$ (resp. $P_{2k+1}=(L^{\frac{2k-1}{2n-2}})_+$   
if $\widetilde{\mathfrak{g}}$ is $D_n^{(1)}$ ) where $L$ is the Lax operator associated to $B_n^{(1)}$ (resp.  
 $D_n^{(1)}$ \cite{DS} \cite{DLZ} and $(P)_+$ denotes the differential part of a pseudo differential operator $P$. But since \cite{DJKM} 
$P_{2k+1}= \sum_{j=1}^{2k+1}p_j\partial_x^j$ it follows that in both cases of affine Kac--Moody Lie algebras of type $B$ or $D$, $H^{(2j+1)}$ has the  expansion in term of Fa\`a di Bruno Polynomials  given by equation (\ref{faexp}).
\endpf
It is here worth to note that also in the case $\widehat{\mathfrak{g}}=B_n^{(1)}$ it is possible to show that  $H^{(2j+1)}$ is completely determined by the  Fa\`a di Bruno Polynomial $h$ without referring to the pseudo differential approach to the Drinfeld--Sokolov hierarchies, it holds indeed.
\begin{prop}\label{BNH} If $\widetilde{\mathfrak{g}}$ is $B_n^{(1)}$ then 
$H^{(2j+1)}$ can be written as
\beq\label{2j1bn}
H^{(2j+1)}=\sum_{i=0}^{2n-1}\left(\sum_{l=0}^{\left[\frac{2j+1-i}{2n}\right]}r^i_l(x)\lambda^l\right)h^{(i)},
\eeq
the coefficients $r_l$ are idependent from $\lambda$ and are completely determined by the fact that $H^{(2j+1)}$ has the Laurent expansion  $H^{(2j+1)}=z^{2j+1}+(\mbox{ negative terms in $z$})$.
\end{prop} 
{\bf Proof} From $H^{(2j+1)}=\left(V_{\Lambda^{2j+1}})^T+v^{(0)},\psi\right)$, $V_{\Lambda^{2j+1}}=T\Lambda^{2j+1}T^{-1}$  the Proposition \ref{gencassol}  and the fact that $((V_{\Lambda^j})_+)_{0,N-1}=0$ follows that 
$$
H^{(2j+1)}=\sum_{i=0}^{2n-1}\left(\sum_{l=0}^{\left[\frac{2j+1}{2n}\right]}r^i_l(x)\lambda^l\right)h^{(i)}.
$$
Now since the maximal power of $H^{(2j+1)}$ is $z^{2j+1}$, it is easy to show by induction that $r^i_l=0$ if $l> \left[\frac{2j+1-i}{2n}\right]$ i.e., that equation (\ref{2j1bn}) holds.\par 
Now since $i$ runs from $0$ to $2n-1$ there exists exact an odd number   say $2k+1$,  $0\leq k\leq n-1$ such that   $2j-2k$ is divisible by $2n$ we have the number of  the coefficients $r^i_l$ in (\ref{2j1bn}) is 
$$
\left(\frac{j-k}{n}+1\right)(2k+2)+\frac{j-k}{n}(2n-1-2k-1)=2(j-k)+2k=2j+2.
$$
But this is exactly the number of independent condition imposed on those coefficients by th  very form of $H^{(2j+1)}$ as Laurent series: $H^{(2j+1)}=z^{2j+1}+(\mbox{ negative terms in $z$})$.
\endpf


\end{document}